# Regional System Identification and Computer Based Switchable Control of a Nonlinear Hot Air Blower System

Ijaz Hussain, Muhammad Riaz, Muhammad Rehan, and Shakeel Ahmed
*Systems and Control Research Group*
*Pakistan Institute of Engineering and Applied Sciences (PIEAS)*
*Islamabad, Pakistan.*
ijazhussain7979@hotmail.com, mohsinriaz18@gmail.com
rehan@pieas.edu.pk, fac244@pieas.edu.pk

***Abstract*- This paper describes the design and implementation of linear controllers with a switching condition for a nonlinear hot air blower system (HABS) process trainer PT326. The system is interfaced with a computer through a USB based data acquisition module and interfacing circuitry. A calibration equation is implemented through computer in order to convert the amplified output of the sensor to temperature. Overall plant is nonlinear; therefore, system identification is performed in three different regions depending upon the plant input. For these three regions, three linear controllers are designed with closed-loop system having small rise time, settling time and overshoot. Switching of controllers is based on regions defined by plant input. In order to avoid the effect of discontinuity, due to switching of linear controllers, parameters of all linear controllers are taken closer to each other. Finally, discretized controllers along with switching condition are implemented for the plant through computer and practical results are demonstrated.**

## I. Introduction

In process industry, temperature control of different systems is often required. Reactions rates are controlled by heating and cooling the reactants. Power production in reactors is based on temperature control of different processes. For industrial applications, automatic temperature control of furnaces is required for melting, decomposing and studying the physical and chemical properties of substances. Temperature control is one of the vital control variables like flow, level, pressure and motor speed. In industry, a fine control of temperature can be a critical issue with consideration of safety of the equipment. Temperature control systems are generally non-linear in nature. Such systems are controlled by both linear and nonlinear controllers [1-4]. Linear controllers are easy to design but their performance cannot be very good. On the other hand, nonlinear controller may have a good performance but they can be difficult to design. Secondly it may not be always possible to model all the dynamics of a nonlinear system. Adaptive controllers are also useful for temperature control applications but they need higher computations.

Different classical techniques are being adopted for the temperature control applications. In [1-2], modeling and linear controller design techniques are discussed for temperature control. In [3-4], linear robust controller design framework based on linear model is proposed for level and temperature control of a nonlinear three tank system. Problem with such classical techniques is: they use information of linear models along with equilibrium or operating point. So controller performance is better around the equilibrium point rather than the whole region. A simple PI controller can also be solution, which is one of most popular controller in the industry. It performs well for systems like position control [5-6], which are actually less nonlinear in nature than temperature control problems. These PI controllers cannot perform well for all operating points of a nonlinear temperature control system. Adaptive PI controllers can also be used for such situations but they require higher number of computations due to complexity of the adaptive algorithms [7].

An alternative solution for temperature control problem is to use a number of PI controllers for different operating regions. Each PI controller must be tuned in such a way that it gives the required performance objectives in that region. There will be slight increase in computations due to additional switching condition which is normally acceptable because it improves the overall closed-loop performance and computational complexity is less than adaptive algorithms.

In this paper, we consider the control of a hot air blower system PT326 by interfacing it with computer through a USB based data acquisition module 1208fs. The process trainer PT326 consists of heater element, motor element, conditioning, amplifying circuitry, sensor and analog temperature scale. The detail description of this system can be found in its User's Manual [8]. This system has facility of proportional and two step controllers, which have performance limitations. For experimental purposes, we can introduce output delays, actuator and sensor constraints and disturbance. Input and output terminals are also accessible for external control. For interfacing purpose, conversion of voltage signals into required acceptable range for 1208fs is done by the addition of Op-amp based circuitry. Measured output voltage is converted into temperature through a calibration equation which is implemented in the computer. System identification is done by applying step input through computer. Rather than identifying a single linear model, we identified three linear models for different ranges of input. Three PI controllers are designed with suitable performance for these identified models. These PI controllers are discretized and implemented through computer in order to control the temperature of system.

In addition to PI controllers, switching condition is also implemented in order to select the controller for a specific region. Within a region, the proposed feedback control is linear but whenever the region is switched, the controller is switched to other which causes a discontinuity. This type of discontinuity can cause undesirable change in rate of actuator or output. So controller parameters for different regions are selected closer to each other which minimize the effect of discontinuity. It is also sometimes required to control an industrial process with different rates for different regions where controller parameter are very different from each other. So switching control can also be useful there. But in the present work, we are considering a performance based switching controller rather than rate based switching controller. Finally the practical results for temperature output and actuator signal are shown and discussed.

## II. Computer Interfacing and Calibration

Process trainer PT 326 system consists of an arrangement of heater and fan attached with a tunnel. This heater and fan assembly is used to blow hot air through the tunnel. A thermister is used to measure the temperature of the hot air. The amplified sensor output in range of 0 to 14 volts is available for external control application. There are three sensor positions at tunnel for inserting thermister. From these sensor positions, different output delays can be introduced (see details in its User Manual as [8]).

In the present work, we are using first position which has no output delay. An input of 0 to 13 volts can be applied at the plant. This system also has actuator and sensor constraints which are not accorded in the present work. Disturbances can also be introduced in the system by changing the angle of throttle.

Hot air blower system is interfaced with computer by using 1208fs USB based data acquisition module and op-amp based circuitry consisting of amplifiers, buffers and subtractor as shown in Figure 1. Amplified thermistor output of 0 to 14 volts is converted into -3 to 8 volts range which is assigned as measured signal. The output range for 1208fs module is between 0 to 4 volts (actuator signal) which is amplified to 0 to 13 volts. Measured signal of -3 to 8 volts is converted into temperature output by using a third order calibration equation.

For calibration, different inputs are applied and corresponding measured voltages at steady state and temperature values are recorded. For best results, averaging of several readings is performed.

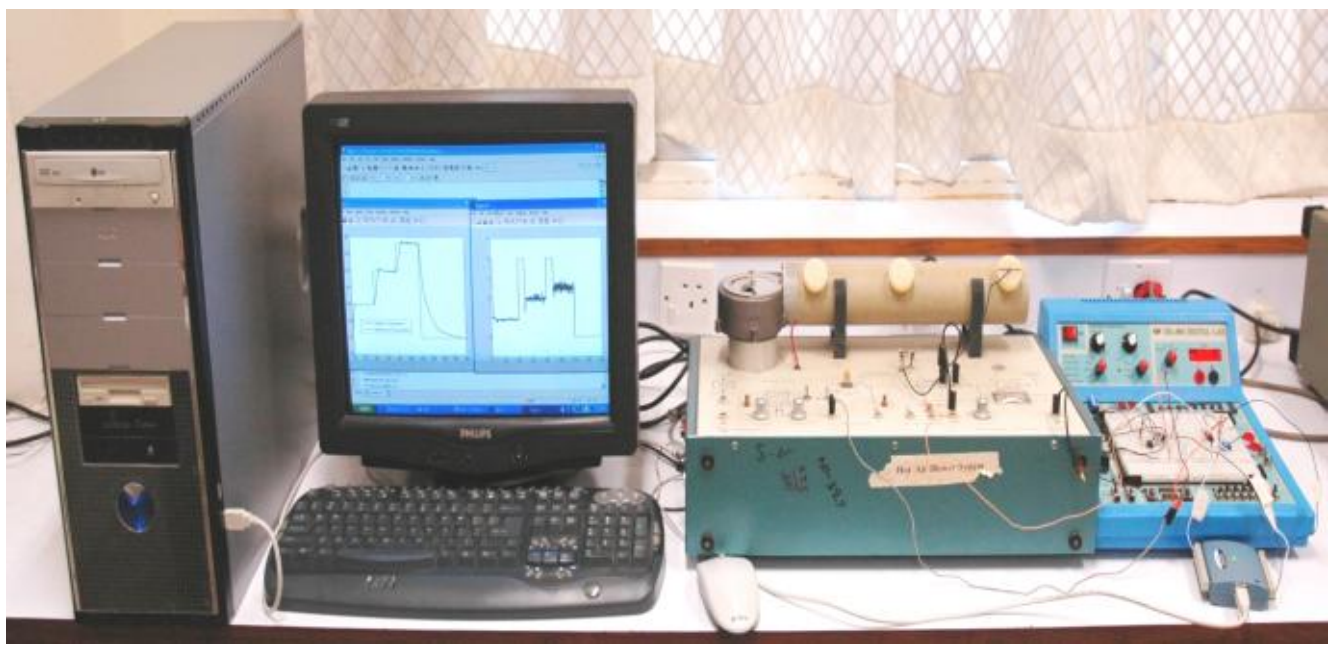

Figure 1. Hot air blower system PT326 interfaced with computer through 1208fs module

Data used for calibration equations is provided in Table 1. In order to build a relation between the measured voltage signal *V* and the corresponding temperature *T*, the following calibration equation is used (see also [1]):

$$T = 0.072\,V^3 - 0.3033\,V^2 + 2.2459\,V + 38.1792 \qquad (1)$$

TABLE 1. Calibration data between Temperature T and Measured Voltage signal V

| Sr. No. | Temperature ($T$ in $^oC$) | Measured Signal ($V$ in Volts) |
|---|---|---|
| 1 | 23 | -3.5299 |
| 2 | 30 | -2.5177 |
| 3 | 40 | 1.2735 |
| 4 | 50 | 4.5324 |
| 5 | 60 | 6.7338 |
| 6 | 70 | 7.5873 |

Figure 2 shows the actual plot between the measured signal *V* and temperature output *T* and plot obtain by third order polynomial of (1).

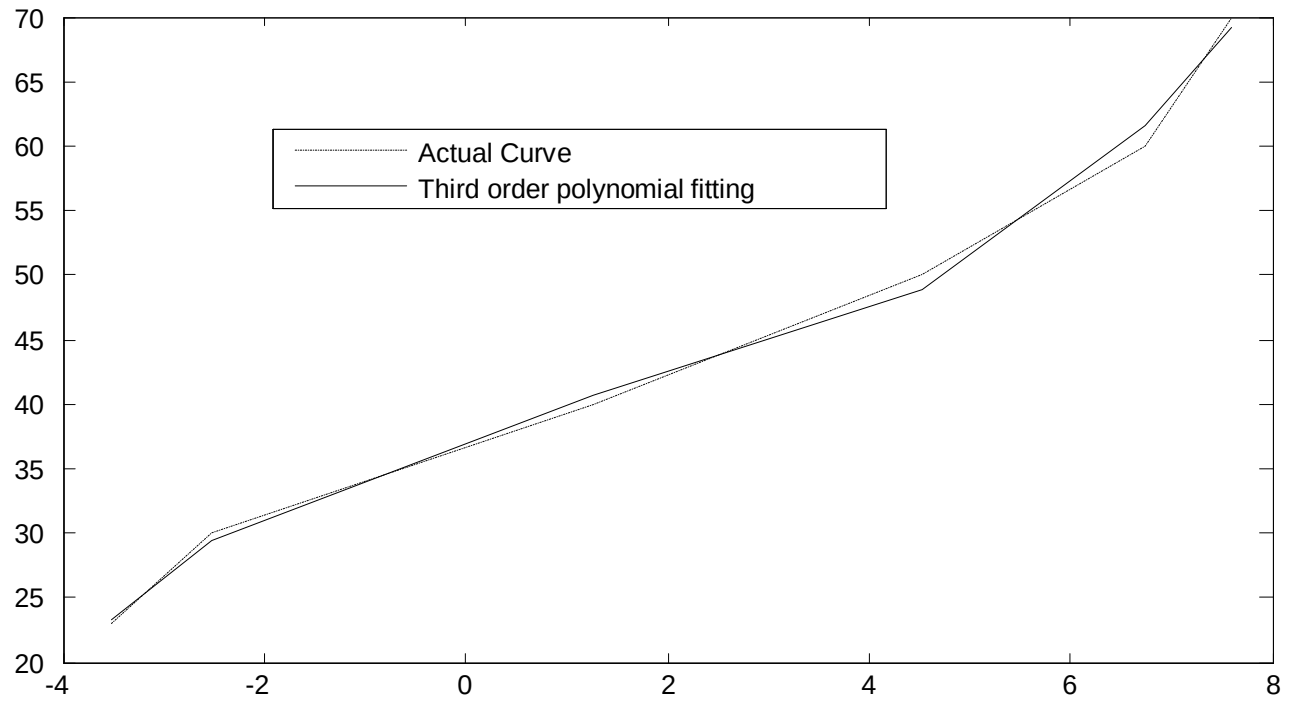

Figure 2. Actual and calibration equation plots between temperature and measured voltage signal

It can be seen in Figure 2 that the error between the actual curve and calibration equation is small. This calibration equation is finally implemented through computer to get the temperature of air blowing through tunnel. We also noted that the plot of Figure 2 between measured signal and temperature is almost linear for 30 to 50°C. Below 30°C and above 50°C, it is slightly nonlinear. So a higher order calibration must be used which justify the third order calibration equation of (1) in our case.

## III. System Identification

The actual model of hot air blower system is nonlinear, which is difficult to find. Secondly, the controller design can also be complex due to nonlinear dynamics [9]. So, we consider the linear identification of hot air blow system in three different regions, depending upon the plant input. As a single linear model cannot describe the whole system for all operating points, so it is better to identify a number of linear models for different regions as plant model varies from one operating region to other. In our case, system identification is performed for three different regions based on input applied through data acquisition module. Output samples are taken at sampling time of 0.1 second. Step response for each region is recorded for a time of one minute. Figure 3 shows the step response for region I when input is changed from 0 to 1 through a computer program. Corresponding temperature changes from 29.6 °C to 39.5 °C.

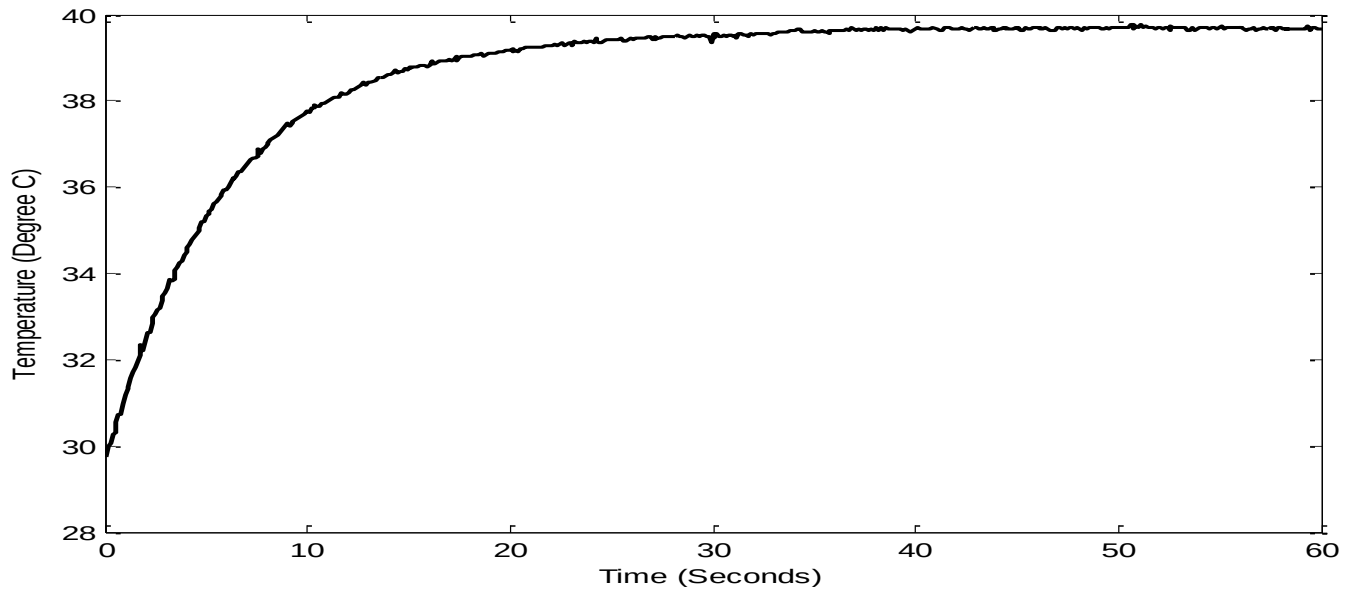

Figure 3. Step response for input of region I

Figures 4 and 5 show the system responses for region II and III, respectively, for a step input of 1-2 and 2-3, respectively. The temperature of air blowing through tunnel changes from 39.5 °C to 48.7 °C and 48.7 °C to 59 °C respectively, for each step. We can clearly see that the plant behavior is changed for each of three step inputs. An increase in rise time and variation in output gain is observed for the plant as we go for region I to III which indicates the nonlinear behavior of hot air blower system.

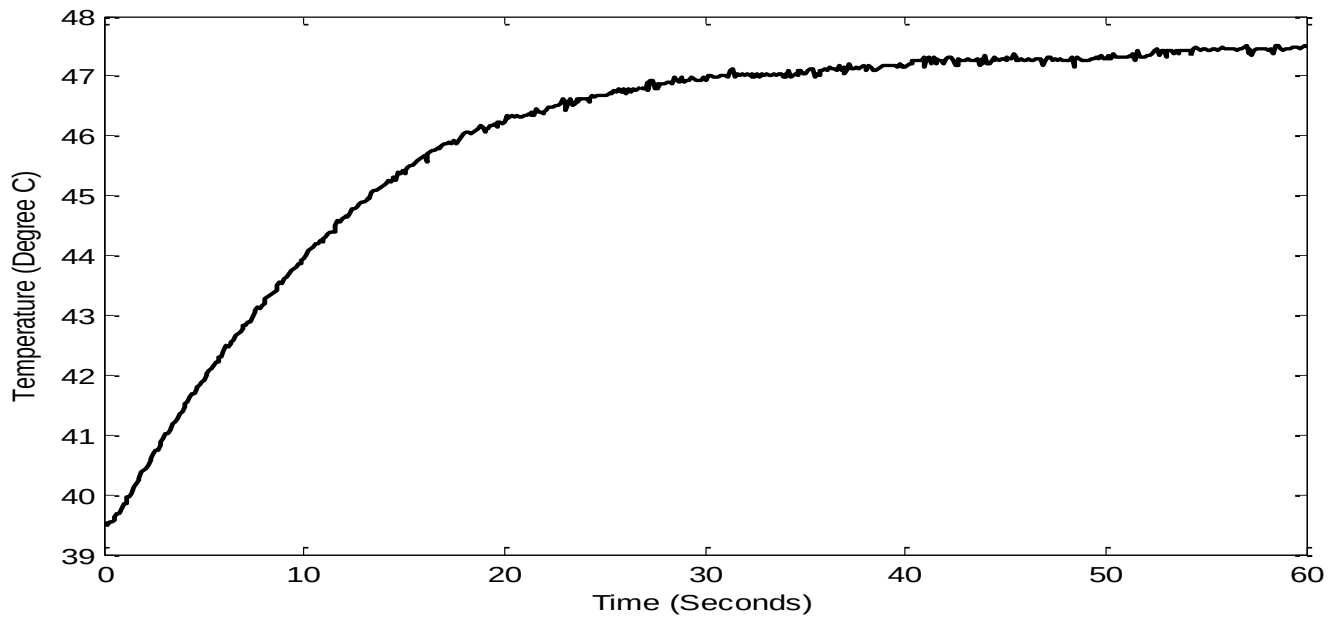

Figure 4. Step response for input of region II

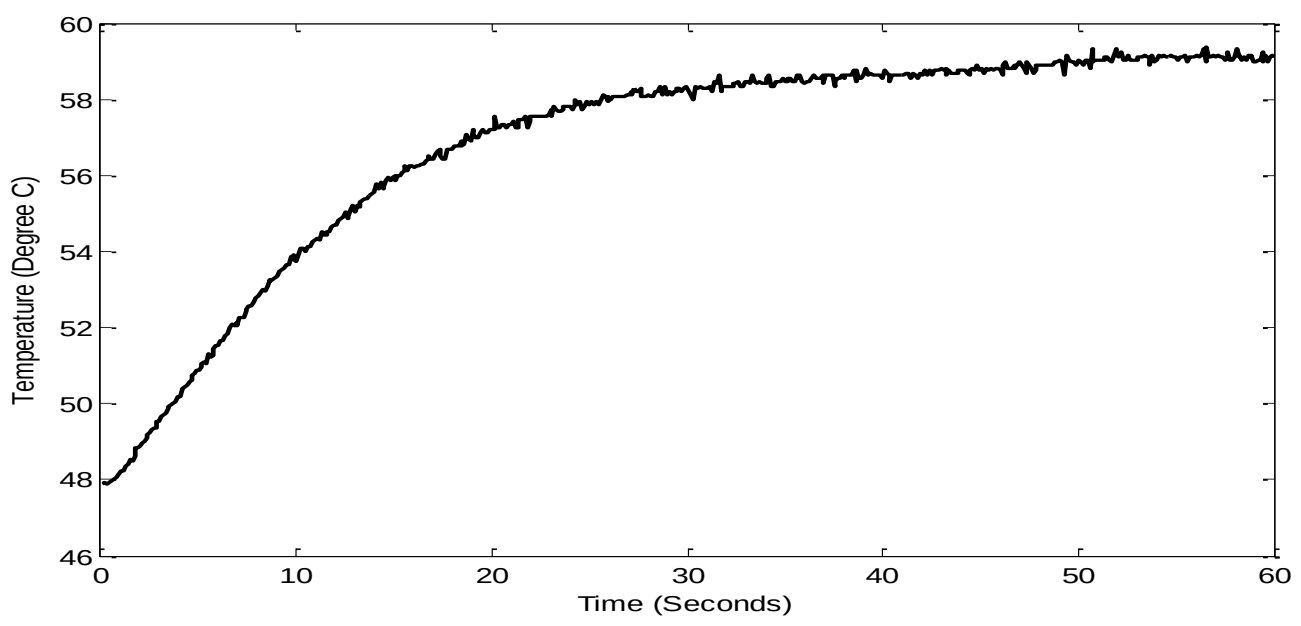

Figure 5. Step response for input of region III

From these step responses, overall identified plant is given by (2):

$$P \sim \begin{cases} G_I(s) = \dfrac{9.5}{6.5s+1} & \forall\ 0 \le u < 1 \\ G_{II}(s) = \dfrac{8}{15s+1} & \forall\ 1 \le u < 2 \\ G_{III}(s) = \dfrac{10}{16s+1} & \forall\ u \ge 2 \end{cases} \quad (2)$$

## IV. Controller Design and Implementation

Controller designed for the model of (2) can be either robust or nonlinear. It is seen in practice that the robust controllers have higher order and increase the computations due to tuning of weights as in [4]. Secondly there can be a trade-off between robustness and performance. So, inspired by standard gain scheduling control schemes [7], we design a nonlinear switchable type control for hot air blower system. For the identified model of (2), three different PI controllers are tuned based on the following objectives:

1. Rise time of the closed-loop system must be less than 4 seconds in order to get fast response.
2. The overshoot for the closed-loop system must be less than 10 percent.
3. The parameters of all PI controllers must be closer to each other in order to decrease the effect of discontinuity due to switching of controllers.

If the parameters of PI controllers differ much from each other then there can be problems due to discontinuity caused by switching of controllers. Hence, we tried to take proportional and integral parameters of all three controllers closer to each other so that the closed-loop response has less affect due to switching of controllers and must look continuous. The overall controller representing all three regions is given by

$$C \sim \begin{cases} K_I(s) = 1.3 + \dfrac{0.11}{s} & \forall\ 0 \le u < 1 \\ K_{II}(s) = 1.5 + \dfrac{0.13}{s} & \forall\ 1 \le u < 2 \\ K_{III}(s) = 1.8 + \dfrac{0.12}{s} & \forall\ u \ge 2 \end{cases} \quad (3)$$

For implementation of (3), we use a check on plant input in the form of switching condition. Whenever plant input changes its region, controller is also changed. In a specific region, control is continuous while discontinuity is observed when region is changed. Hence the overall control is a combination of continuous and discontinuous control. Controller architecture is shown in Figure 6. Controller has two inputs. One is the difference between reference signal and output temperature while other is previous value of actuator signal in order to decide the controller for a specific region. The output of controller is feed into plant through data acquisition module and interfacing circuitry. The output of PT326 is converted into required range by interfacing circuitry and then further converted into temperature by calibration equation which is implemented in computer.

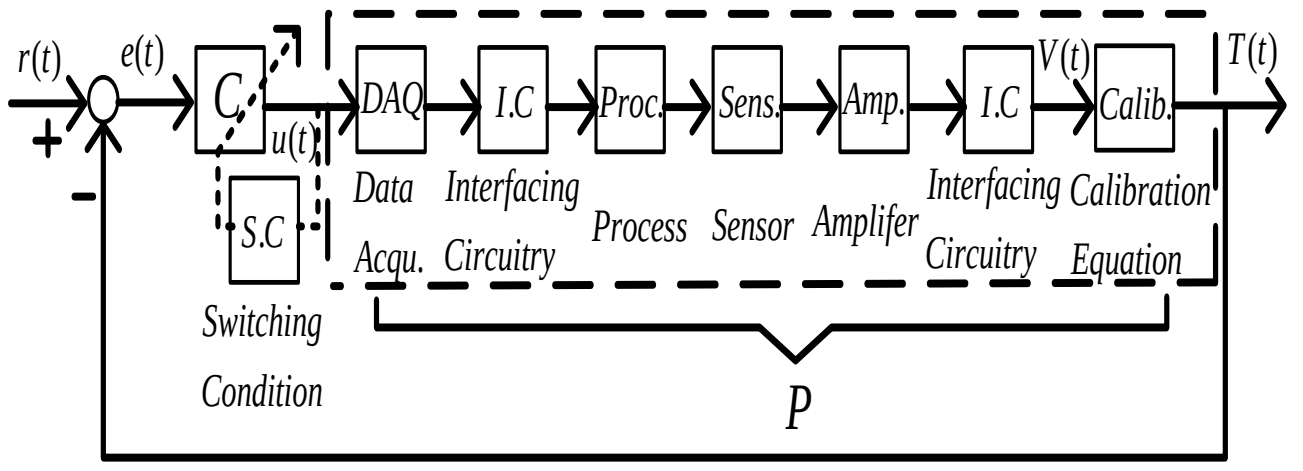

Figure 6. OVERALL SWITCHABLE CONTROL STRUCTURE FOR HOT AIR BLOWER SYSTEM

Switching controllers are discreetized for a sampling time of 0.1 second using bilinear approximation and implemented through a computer program. Figures 7 and 8 show the output response of the overall closed-loop system and actuator signal respectively. The output response is tracking the reference for all three regions. Discontinuity due to switching is so small that it is not observed in both output and actuator signals. Due to actuator saturation, the response is not fast in region II and III as compared to region I for rising step. A little overshoot is observed in closed-loop response due to nonlinear dynamics of plant and actuator saturation.

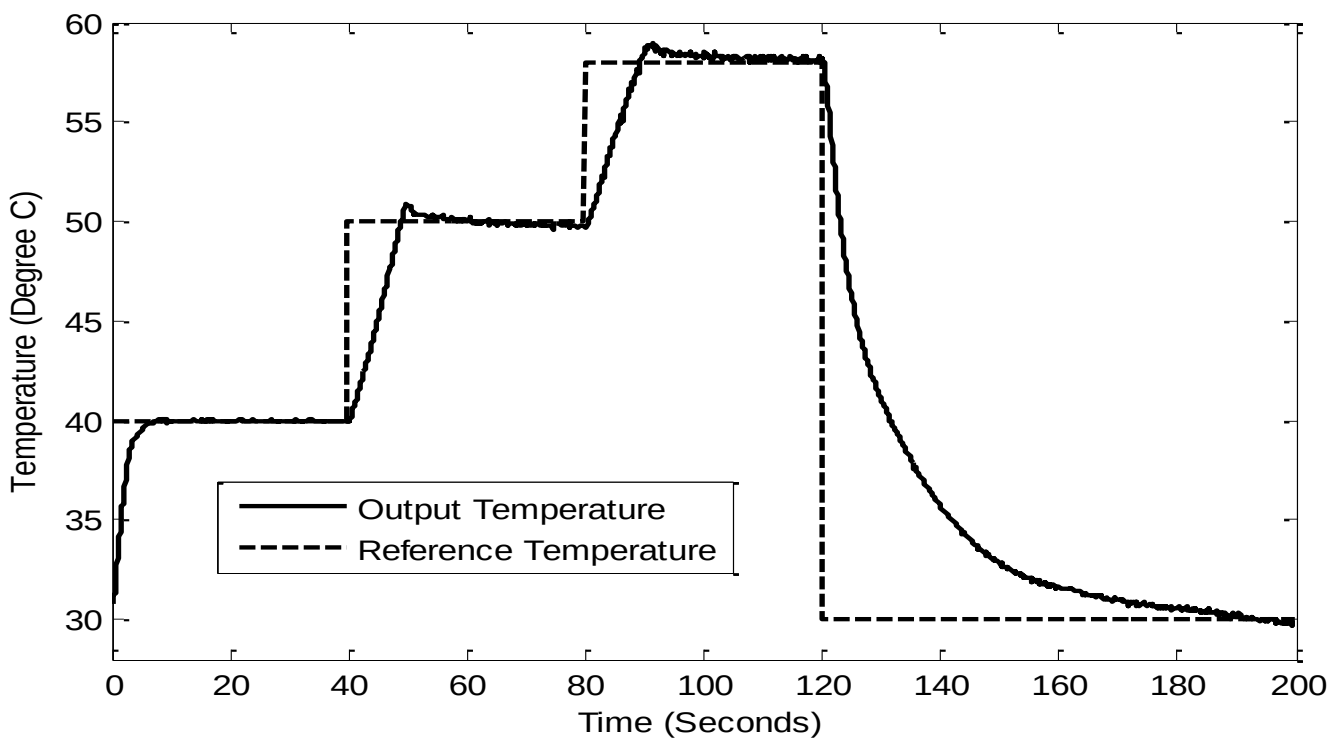

Figure 7. CLOSED-LOOP RESPONSE FOR SWITCHABLE CONTROL OF HOT AIR BLOWER SYSTEM

Higher frequency variations are observed in actuator signal. These variations are found to be dominant for higher values of actuator. Such variations are also present at output response as well. At actuator signal, these high frequency variations are amplified due to controller gains. There can be two reasons for the increase in amplitude of high frequency variations as actuator signal settles at higher value. One is thermal noise which generally increases due to increase in temperature. Secondly for the present scenario, the proportional gain increases as we go from region I to III.

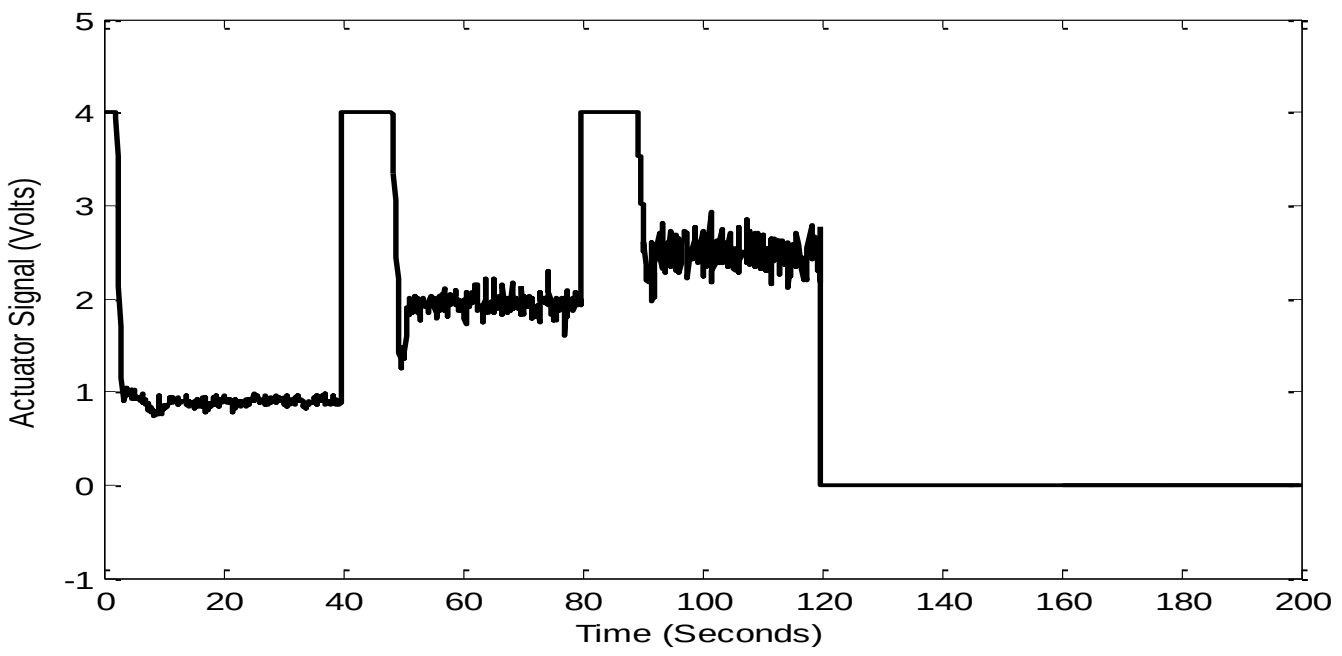

Figure 8. ACTUATOR RESPONSE FOR SWITCHABLE CONTROL OF HOT AIR BLOWER SYSTEM

## V. CONCLUSION

This paper described the computer interfacing, calibration, system identification for different regions and design and implementation of different PI controllers with switching condition for temperature control of a nonlinear hot air blower system PT326. The plant was interfaced with computer using USB based data acquisition module 1208fs and op-amp based electronic circuitry. Temperature output was obtained by implementing a third order calibration equation between temperature and measured voltage signal. System identification was performed for three different regions of input signals. Three PI controllers with slightly varied parameters were designed. These PI controllers were discreetized and implemented along with calibration equation and switching condition through a computer program. The overall closed-loop response was tracking the reference. Actuator signal was found to have an increase in amplitude for the high frequency variations as temperature was increased. This methodology can be useful for performance oriented temperature based process control.